\documentclass[aps,prb,twocolumn,groupedaddress,showpacs]{revtex4}

\usepackage{graphicx}

\begin{document}


\title{Spontaneous exciton condensation in 1\textit{T}-TiSe$_2$: a BCS-like approach}

\author{C. Monney}
\email{claude.monney@unine.ch}
\author{H. Cercellier}
\author{F. Clerc}
\author{C. Battaglia}
\author{E.F. Schwier}
\author{C. Didiot}
\author{M.G. Garnier}
\author{H. Beck}
\author{P. Aebi}%

\affiliation{%
Institut de Physique, Universit\'e de Neuch\^atel, CH-2000 Neuch\^atel, Switzerland}%

\author{H. Berger and L. Forr\'o}
\affiliation{Institut de Physique de la Mati\`ere Complexe, EPFL, CH-1015 Lausanne, Switzerland
}%

\author{L. Patthey}
\affiliation{Swiss Light Source, Paul Scherrer Institute, CH-5232 Villigen, Switzerland}

\date{\today}

\begin{abstract}
Recently strong evidence has been found in favor of a BCS-like condensation of excitons in 1\textit{T}-TiSe$_2$. Theoretical photoemission intensity maps have been generated by the spectral function calculated within the excitonic condensate phase model and set against experimental angle-resolved photoemission spectroscopy data. Here, the calculations in the framework of this model are presented in detail. They represent an extension of the original excitonic insulator phase model of J\'erome \textit{et al.} [Phys. Rev. {\bf 158}, 462 (1967)] to three dimensional and anisotropic band dispersions. A detailed analysis of its properties and further comparison with experiment are also discussed.
\end{abstract}


\maketitle

\section{Introduction}

In the early 1960s, a new insulating phase was predicted to possibly exist at low temperature in solids having small energy gaps. J\'erome \textit{et al.}\cite{JeromeBasis} published an extended study of this phase developing a BCS-like theory of its ground state. However, at that time an experimental realization of this phase was missing.

The excitonic insulator phase may occur in a semi-metallic or semiconducting system exhibiting a small (negative respectively positive) gap. Indeed, for a low carrier density, the Coulomb interaction is weakly screened, allowing therefore bound states of holes and electrons, called excitons, to build up in the system. If the binding energy $E_B$ of such pairs is larger than the gap $E_G$, the energy to create an exciton becomes negative, so that the ground state of the normal phase becomes unstable with respect to the spontaneous formation of excitons. According to J\'erome \textit{et al.}\cite{JeromeBasis}, at low temperature, these excitons may condense into a macroscopic coherent state in a manner similar to Cooper pairs in conventional BCS superconductors. Kohn\cite{KohnCDW} argued that exciton condensation may lead to the formation of charge density waves (CDW) of purely electronic origin (neglecting any lattice distortion), characterized by an order parameter.

1\textit{T}-TiSe$_2$ is a layered transition-metal dichalcogenide exhibiting a commensurate (2x2x2) CDW\cite{DiSalvoSupLatt} accompanied by a periodic lattice distortion below the transition temperature of $T_c\cong 200$ K. The origin of its CDW phase was controversial for a long time. Different scenarios have been proposed \cite{nesting,antiferro}, the best candidates being a band Jahn-Teller effect \cite{HughesBJT} and the excitonic insulator phase. In 2006, superconductivity has been discovered in TiSe$_2$ upon Cu intercalation, providing a renewed interest in this system\cite{MorosanSupraCuTiSe2}. Furthermore superconductivity also occurs for the pure compound under pressure \cite{Sipos}.
Recently, angle-resolved photoemission spectroscopy (ARPES) data on 1\textit{T}-TiSe$_2$ were presented \cite{CercellierPRL}. Theoretical photoemission intensity maps generated by the spectral function computed within the excitonic condensate phase model gave strong evidence for excitonic condensation in this material. To our knowledge, 1\textit{T}-TiSe$_2$ is the only presently known candidate for a low temperature phase transition to the excitonic condensate state without the influence of any external parameters other than temperature. Indeed, as pressure is increased above 6 kbar on TmSe$_{0.45}$Te$_{0.55}$ samples (allowing to control the gap size and thus the energy necessary to create excitons), a transition to an insulating phase happens, whose origin can also be explained with exciton condensation\cite{WachterPRB}. In this context Bronold and Fehske proposed an effective model for calculating the phase boundary of a pressure-induced excitonic insulator, in the spirit of a crossover from a Bose-Einstein to a BCS condensate\cite{BronoldPRB}.

In this work, we present the theory from which we compute the spectral function used to describe photo\-emission on TiSe$_2$ and provide further support for the exciton phase scenario. In section \ref{sec_model}, we extend the model worked out by J\'erome \textit{et al.}\cite{JeromeBasis} for one dimension to three dimensional and anisotropic band dispersions. The Green's functions of the different bands are derived here. Section \ref{sec_results} first introduces the spectral function and its relation to photoemission. Then spectral weights and positions of the different bands are analyzed within this model. 
These theoretical results are compared to ARPES data of TiSe$_2$. Finally, the chemical potential and discrepancies with density functional theory (DFT) are discussed before we conclude in section \ref{sec_conclusion}.

\section{The excitonic condensate model\label{sec_model}}

In this section, we present the model from which the spectral function describing photoemission on TiSe$_2$ has been computed. J\'erome, Rice and Kohn \cite{JeromeBasis} have already treated in detail the case of a one-dimensional excitonic insulator. In their work, they consider a single valence band and a single conduction band, both isotropic. However, for comparison with real experiment on the electronic structure of TiSe$_2$ an extension of the model to three dimensions with anisotropic band dispersions is required.

\subsection{Description of the model}

\begin{figure}
\centering
\includegraphics[width=7.5cm]{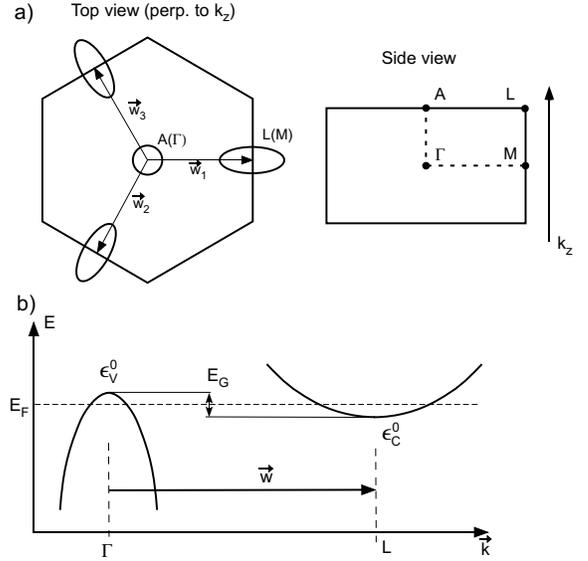}
\caption{\label{fig_schemebands} Schematic picture of the 1\textit{T}-TiSe$_2$ bands considered in this model (near the Fermi energy $E_F$). (a) Top view of the BZ (perpendicular to $k_z$). The Fermi surface has a hole pocket at $\Gamma$ and three symmetry equivalent electron pockets at $L$, separated from $\Gamma$ by the spanning vectors $\vec{w}_i$, $i=1,2,3$. 
The side view of the BZ helps to situate the high symmetry points. (b) Schematic cut along the $\Gamma L$ direction, showing the dispersions of the valence band (at $\Gamma$) and one conduction band (at $L$).
}
\end{figure}

The Hamiltonian of the model is composed of a one-electron part $H_0$ and a Coulomb interaction part $W$. The one-electron part contains the dispersions of a single valence band $\epsilon_v(\vec{k})$ and of three conduction bands $\epsilon_c^i(\vec{k})$ ($i=1,2,3$)
\begin{displaymath}
H_0=\sum_{\vec{k}}\epsilon_v(\vec{k})a^\dagger(\vec{k})a(\vec{k})+\sum_{\vec{k},i}\epsilon_c^i(\vec{k}+\vec{w}_i)b_i^\dagger(\vec{k})b_i(\vec{k}).
\end{displaymath}
Here $a^\dagger(\vec{k})$ and $b_i^\dagger(\vec{k})$ are operators creating electrons with wave vector $\vec{k}$  in the valence band and with wave vector $\vec{k}+\vec{w}_i$ in the conduction band labelled $i$, respectively. In the case of TiSe$_2$, we consider the valence band (mainly of Se 4p character) giving rise to a hole pocket centered at $\Gamma$ and three conduction bands (mainly of Ti 3d character), equivalent by symmetry, giving rise to electron pockets centered at the different L points of the Brillouin zone (BZ) (see Fig. \ref{fig_schemebands} for a sketch of high symmetry points in the BZ). The $\Gamma$ point is separated from the $L$ points by the three spanning vectors $\vec{w}_i=\Gamma L$. The band dispersions have been chosen of the form
\begin{eqnarray}
\epsilon_v(\vec{k})&=&\hbar^2\frac{k_x^2+k_y^2}{2m_v}+t_v\cos\left(\frac{2\pi k_z}{2k_{\Gamma A}}\right)+\epsilon_v^0,\nonumber\\
\epsilon_c^i(\vec{k})&=&\frac{\hbar^2}{2m_L}\left((\vec{k}-\vec{w}_i)\cdot \vec{e}_{i\parallel}\right)^2+\frac{\hbar^2}{2m_S}\left((\vec{k}-\vec{w}_i)\cdot \vec{e}_{i\perp}\right)^2\nonumber\\&&+t_c\cos\left(\frac{2\pi (k_z-w_{iz})}{2k_{\Gamma A}}\right)+\epsilon_c^0,\nonumber
\end{eqnarray}
which describe well the bands near their extrema as they are measured in ARPES experiment\cite{CercellierPRL}. 
The unit vectors $\vec{e}_{i\parallel}$ and $\vec{e}_{i\perp}$, pointing along the long and short axis of the ellipses, respectively, form a local in-plane basis  for the electron pockets at the different $L$ points. 
Thus, $\vec{e}_{i\parallel}=\vec{w}_{i\parallel}/||w_{i\parallel}||$ where $\vec{w}_{i\parallel}=(w_{ix},w_{iy},0)$ and $\vec{e}_{i\perp}=\vec{w}_{i\perp}/||w_{i\perp}||$ where $\vec{w}_{i\perp}=(0,0,1)\times \vec{w}_i$. 
The $m_v$, $m_L$ and $m_S$ are the effective masses of the valence band holes and of the conduction band electrons along the long and short axis of the electron pockets, respectively. 
The hopping parameters $t_v$ and $t_c$ represent the amplitudes of the dispersions perpendicular to the surface and $k_{\Gamma A}$ is the distance in reciprocal space between $\Gamma$ and the A point. Parameters $\epsilon_v^0$ and $\epsilon_c^0$ are the band extrema of the bands.

The interaction part $W$ contains only the direct Coulomb interaction between electrons in the valence and the conduction bands
\begin{displaymath}
W=\sum_{\vec{q},i}\rho_a(\vec{q})V_c(\vec{q})\rho_{b,i}^\dagger(\vec{q},\vec{w}_i)\nonumber
\end{displaymath}
where we have introduced partial electron density operators
\begin{displaymath}
\rho_a(\vec{q})=\sum_{\vec{k}}a^\dagger(\vec{k}+\vec{q})a(\vec{k}),\:\:\rho_{b,i}(\vec{q},\vec{w}_i)=\sum_{\vec{k}}b_i^\dagger(\vec{k}+\vec{q})b_i(\vec{k}).\nonumber
\end{displaymath}
The above sums run over the first BZ. Finally the Coulomb potential reads
\begin{displaymath}
V_c(\vec{q})=\frac{4\pi e^2}{\epsilon(\vec{q}) q^2},\nonumber
\end{displaymath}
with $\epsilon$ being the dielectric function of the material.

In fact for 1\textit{T}-TiSe$_2$ there are three (4$p$-derived and Se related valence) bands at $\Gamma$ near the Fermi energy. However, within a minimal model, we include only the highest lying valence band (as the valence band) in the calculations. Finally, the chemical potential is not explicitly included in the model but it will be nonetheless discussed at the end of section \ref{sec_results}.

\subsection{Exciton physics}

In this paragraph, we introduce the formulation of the exciton physics in a similar way to Babichenko and Kiselev \cite{BabichenkoBasis}. The aim of this discussion is to give a better insight
into the concept of the excitons although it is not essential for the understanding of this paper. Furthermore the formula of this paragraph are not used for the results presented in this article. We start from the field operator $\psi_a(\vec{r})=\sum_{\vec{k}}{\rm e}^{i\vec{k}\cdot\vec{r}}a(\vec{k})$ which creates a hole in the valence band at position $\vec{r}$ and $\psi_{b_i}^\dagger(\vec{r})=\sum_{\vec{k}}{\rm e}^{-i(\vec{k}+\vec{w}_i)\cdot\vec{r}}b_i^\dagger(\vec{k})$ which creates an electron in the conduction band $i$ at position $\vec{r}$. From these two entities, we construct the exciton creation operator
\begin{eqnarray}
A^\dagger(\vec{r},\vec{r}\:',\vec{w}_i)&=&\psi_{b_i}^\dagger(\vec{r})\psi_a(\vec{r}')\nonumber\\
&=&\sum_{\vec{k}_1,\vec{k}_2}{\rm e}^{-i(\vec{k}_1+\vec{w}_i)\cdot\vec{r}}{\rm e}^{i\vec{k}_2\cdot\vec{r}'}b_i^\dagger(\vec{k}_1)a(\vec{k}_2).\nonumber
\end{eqnarray}
It is more appropriate to describe the exciton in terms of relative $\vec{u}$ and center of mass $\vec{R}$ coordinates. Due to the anisotropy of the electron pockets at $L$, we need to distinguish the in-plane coordinates parallel and perpendicular to the long axis of the ellipses, $\vec{w}_{i\parallel}$. For simplicity, we admit that $\vec{w}_{i\parallel}$ is parallel to the $x$-axis (for the other ellipses it is possible to generalize the following arguments using the $\vec{e}_{i\parallel}$ and $\vec{e}_{i\perp}$ unit vectors). 
Then we write $\vec{u}=\vec{r}-\vec{r}\:'$ and $R_\alpha=m_\alpha' r_\alpha'/M_\alpha+m_\alpha r_\alpha/M_\alpha$ with $M_\alpha=m_\alpha'+m_\alpha$ for $\alpha=x,y$. In terms of the previously defined masses we have
$m_\alpha'=m_v$, since the hole pocket at $\Gamma$ is isotropic, and $m_x=m_L$ and $m_y=m_S$.
Then, the exciton creation operator may be redefined as
\begin{eqnarray}
A^\dagger(\vec{R},\vec{u},\vec{w}_i)&=&\sum_{\vec{Q},\vec{p}}{\rm e}^{-i(\vec{Q}+\vec{w}_i)\cdot\vec{R}} \  {\rm e}^{-i\vec{p}\cdot\vec{u}-i\sum_\alpha\frac{m_a'}{M_\alpha}w_{i\alpha}u_\alpha}\nonumber\\& &\times\: b_i^\dagger\left(p_\alpha+\frac{m_\alpha}{M_\alpha}Q_\alpha\right)a\left(p_\alpha-\frac{m_\alpha'}{M_\alpha}Q_\alpha\right)\nonumber
\end{eqnarray}
using the notation $a(k_\alpha)$ instead of $a((k_x,k_y))\equiv a(\vec{k})$. The center of mass momentum $\vec{Q}=\vec{k}_1-\vec{k}_2$ and the relative momentum $p_\alpha=m_\alpha k_{1,\alpha}/M_\alpha+m_\alpha'k_{2,\alpha}/M_\alpha$ arise naturally.

At this point, we can expand the operator $b^\dagger a$ in terms of the excitonic creation operator $A^\dagger$ in reciprocal space
\begin{eqnarray}\label{eqn_barecip}
b_i^\dagger\left(p_\alpha+\frac{m_\alpha}{M_\alpha}Q_\alpha\right)a\left(p_\alpha-\frac{m_\alpha'}{M_\alpha}Q_\alpha\right)=\nonumber\\\sum_{\lambda}\phi_\lambda^*(\vec{p},\vec{w}_i)A_\lambda^\dagger(\vec{Q},\vec{w}_i)
\end{eqnarray}
where the coefficients appearing on the right hand side are the eigenfunctions of the hydrogen atom. In other words, the operator $A_\lambda^\dagger (\vec{Q},\vec{w}_i)$ creates an exciton having a center of mass momentum $\vec{Q}$. The electron-hole bound state is described by the hydrogen state $\phi_\lambda$ having the energy $E_\lambda=\mu e^4/8\epsilon^2\lambda^2$, $\epsilon$ being the dielectric constant and $1/\mu=\sum_\alpha 1/2\mu_\alpha$ being the reduced mass with $1/\mu_\alpha=1/m_\alpha+1/m_\alpha'$. According to Babichenko and Kiselev \cite{BabichenkoBasis} (and generalizing to anisotropic conduction bands), this hydrogen state obeys to
\begin{displaymath}
\left(\sum_\alpha\frac{p_\alpha^2}{2\mu_\alpha}+E_\lambda\right)\phi_\lambda^*(\vec{p},\vec{w}_i)=\sum_{\vec{p}\:'}V_c(\vec{p}-\vec{p}\:')\phi_\lambda^*(\vec{p}\:',\vec{w}_i).\nonumber
\end{displaymath}
Due to orthogonality of the hydrogen wave functions, the relation (\ref{eqn_barecip}) can be inverted to 
\begin{eqnarray}
A_\lambda^\dagger(\vec{Q},\vec{w}_i)&=&\nonumber\\\sum_{\vec{p}}&\phi_\lambda^*(\vec{p},\vec{w}_i)&b_i^\dagger\left(p_\alpha+\frac{m_\alpha}{M_\alpha}Q_\alpha\right)a\left(p_\alpha-\frac{m_\alpha'}{M_\alpha}Q_\alpha\right).\nonumber
\end{eqnarray}
After this parenthesis which gave detail about the physics of excitons in the framework of our model, we now compute the equations of motion for annihilation operators.

\subsection{Equations of motion for the Green's functions}

With the help of the Hamilonian $H=H_0+W$, we compute the equation of motion for our electron annihilation operators
\begin{eqnarray}
\lefteqn{i\frac{\partial}{\partial t}a(\vec{p},t)= [a(\vec{p},t),H] =\epsilon_v(\vec{p})a(\vec{p},t)}\nonumber\\&&+\sum_{\vec{q},\vec{k},i}V_c(\vec{q})a(\vec{p}+\vec{q},t)b_i^\dagger(\vec{k},t)b_i(\vec{k}-\vec{q},t), \nonumber\\
\lefteqn{i\frac{\partial}{\partial t}b_i(\vec{p},t)= [b_i(\vec{p},t),H]=\epsilon_c^i(\vec{p}+\vec{w}_i)b_i(\vec{p},t)}\nonumber\\&&+\sum_{\vec{q},\vec{k}}V_c(\vec{q})a^\dagger(\vec{k}+\vec{q},t)a(\vec{k},t)b_i(\vec{p}+\vec{q},t) \label{eqn_MotionOps}.
\end{eqnarray}
We now introduce Green's functions for the valence and the conduction bands
\begin{eqnarray}
G_v(\vec{k},t,t')&=&(-i)\langle T a(\vec{k},t)a^\dagger(\vec{k},t')\rangle,\nonumber\\
G_c^i(\vec{k},t,t')&=&(-i)\langle T b_i(\vec{k},t)b_i^\dagger(\vec{k},t'),\rangle\nonumber
\end{eqnarray}
where we used the time ordering operator $T$.
Their equations of motion are derived directly from equations (\ref{eqn_MotionOps})
\begin{eqnarray}
\lefteqn{\left(i\frac{\partial}{\partial t}-\epsilon_v(\vec{p})\right)G_v(\vec{p},t,t')=\delta(t-t')}\nonumber\\&&-i\sum_{\vec{q},\vec{k},i}V_c(\vec{q})\langle T a(\vec{p}+\vec{q},t)b_i^\dagger(\vec{k},t)b_i(\vec{k}-\vec{q},t)a^\dagger(\vec{p},t')\rangle, \nonumber\\
\lefteqn{\left(i\frac{\partial}{\partial t}-\epsilon_c^i(\vec{p}+\vec{w}_i)\right)G_c^i(\vec{p},t,t')=\delta(t-t')}\nonumber\\&&-i\sum_{\vec{q},\vec{k}}V_c(\vec{q})\langle T a^\dagger(\vec{k}+\vec{q},t)a(\vec{k},t)b_i(\vec{p}+\vec{q},t)b_i^\dagger(\vec{p},t')\rangle .\nonumber
\end{eqnarray}
Using Wick's theorem we simplify the four-operator averages $\langle ...\rangle$ by neglecting correlations, i.e. keeping only the lowest order terms. The calculation is similar for both Green's functions. We get three two-operator contributions, namely
\begin{eqnarray}
\langle Ta(\vec{p}+\vec{q},t)b_i^\dagger(\vec{k},t)b_i(\vec{k}-\vec{q},t)a^\dagger(\vec{p},t')\rangle=\nonumber\\
\langle a(\vec{p}+\vec{q},t)b_i^\dagger(\vec{k},t)\rangle\langle Tb_i(\vec{k}-\vec{q},t)a^\dagger(\vec{p},t')\rangle\nonumber\\
-\langle a(\vec{p}+\vec{q},t)b_i(\vec{k}-\vec{q},t)\rangle\langle Tb_i^\dagger(\vec{k},t)a^\dagger(\vec{p},t')\rangle\nonumber\\
-\langle Ta(\vec{p}+\vec{q},t)a^\dagger(\vec{p},t')\rangle\langle b_i(\vec{k}-\vec{q},t)b_i^\dagger(\vec{k},t)\rangle,\nonumber
\end{eqnarray}
out of which only the first one remains (the second one is zero and the last one is a Hartree term which we consider as already included in the measured dispersions) so that the corresponding equations of motion become
\begin{eqnarray}\label{eqn_valGreendev}
\lefteqn{\left(i\frac{\partial}{\partial t}-\epsilon_v(\vec{p})\right)G_v(\vec{p},t,t')\approx\delta(t-t')}\nonumber\\&&+i\sum_{\vec{q},\vec{k},i}V_c(\vec{q})\langle b_i^\dagger(\vec{k},t)a(\vec{p}+\vec{q},t)\rangle\langle Tb_i(\vec{k}-\vec{q},t)a^\dagger(\vec{p},t')\rangle, \nonumber\\
\lefteqn{\left(i\frac{\partial}{\partial t}-\epsilon_c^i(\vec{p}+\vec{w}_i)\right)G_c^i(\vec{p},t,t')\approx\delta(t-t')}\nonumber\\&&-i\sum_{\vec{q},\vec{k}}V_c(\vec{q})\langle b_i(\vec{p}+\vec{q},t)a^\dagger(\vec{k}+\vec{q},t)\rangle\langle Ta(\vec{k},t)b_i^\dagger(\vec{p},t')\rangle .\nonumber\\
\end{eqnarray}
At this point, to go further we need to introduce the concept of the condensate phase.

\subsection{Condensate phase}

When the energy gap is smaller than the exciton binding energy, the energy necessary to create an exciton becomes negative and the normal ground state becomes unstable towards the spontaneous formation of excitons. Once temperature is low enough, these excitons may condense into a macroscopic coherent state analogous to that of Cooper pairs in the BCS theory of superconductivity.

The first average on the right hand side of equations (\ref{eqn_valGreendev}), for $\vec{k}=\vec{p}+\vec{q}$, can be expressed by the excitonic creation operator (equation (\ref{eqn_barecip}))
\begin{eqnarray}
\langle b_i^\dagger(\vec{p}+\vec{q},t)a(\vec{p}+\vec{q},t)\rangle&=&\sum_{\lambda}\phi_\lambda^*(0,\vec{w}_i)\langle A_\lambda^\dagger(\vec{p}+\vec{q},\vec{w}_i)\rangle\nonumber\\&\approx&\phi_0^*(0,\vec{w}_i)\langle A_0^\dagger(\vec{p}+\vec{q},\vec{w}_i)\rangle.\nonumber
\end{eqnarray}
At sufficiently low temperature, only the lowest lying excitonic level is populated. 

By analogy with the BCS theory, we can identify the average in this last equation with anomalous Green's functions, after appropriate variable substitutions. These new functions are defined as follows 
\begin{eqnarray}
F_i(\vec{k},t,t')&=&(-i)\langle T b_i(\vec{k},t)a^\dagger(\vec{k},t')\rangle,\label{eqn_anomGreen}\nonumber\\
F_i^\dagger(\vec{k},t,t')&=&(-i)\langle T a(\vec{k},t)b_i^\dagger(\vec{k},t')\rangle.\label{eqn_anomGreenCc}
\end{eqnarray}
They describe the scattering of a valence electron into the conduction band or inversely. Pushing further the analogy, we introduce the order parameter $\Delta_i$ describing the condensate
\begin{eqnarray}\label{eqn_orderparam}
\Delta_i(\vec{p})&=&-i\sum_{\vec{q}}V_c(\vec{q})F_i^\dagger(\vec{p}+\vec{q},t,t)\rangle\nonumber\\
&=&\sum_{\vec{q}}V_c(\vec{q})\langle b_i^\dagger(\vec{p}+\vec{q},t)a(\vec{p}+\vec{q},t)\rangle\\
&\approx&\sum_{\vec{q}}V_c(\vec{q})\phi_0^*(0,\vec{w}_i)\langle A_0^\dagger(\vec{p}+\vec{q},\vec{w}_i)\rangle \nonumber
\end{eqnarray}
(here the anomalous Green's function definition (\ref{eqn_anomGreenCc}) and equation (\ref{eqn_barecip}) have been used). It quantifies the intensity of exciton formation between the valence band and the conduction band labelled $i$. Moreover, it characterizes the state of the system in the sense that, when the order parameter is different than zero, exciton condensation drives the system into the CDW phase (see section \ref{sec_results} for further discussion).

\subsection{The Green's function of the valence band}

With the help of these new elements, we can go back to equation (\ref{eqn_valGreendev}). In the particular case of the Green's function of the valence band, we obtain
\begin{eqnarray}\label{eqn_valencetime}
\left(i\frac{\partial}{\partial t}-\epsilon_v(\vec{p})\right)G_v(\vec{p},t,t')=&\delta(t-t')&\nonumber\\-\sum_{i}&\Delta_i(\vec{p}+\vec{q})& F_i(\vec{p},t,t'). 
\end{eqnarray}
In order to solve this equation for $G_v$, we need to find a similar expression for the anomalous Green's function by computing its equation of motion. This procedure results in the following relation
\begin{eqnarray}\label{eqn_anomaltime}
\left(i\frac{\partial}{\partial t}-\epsilon_c^i(\vec{p}+\vec{w}_i)\right)F_i(\vec{p},t,t')=-\Delta_i(\vec{p})G_v(\vec{p},t,t') 
\end{eqnarray}
where we have again identified the order parameter.

Converting the time-dependence into a (imaginary) frequency $z$ dependence with a Fourier transform allows us to solve the system of equations given by (\ref{eqn_valencetime}) and (\ref{eqn_anomaltime}) for the Green's function of the valence band
\begin{eqnarray} \label{eqn_valencefreq}
G_v(\vec{p},z)=\left(z-\epsilon_v(\vec{p})-\sum_i\frac{|\Delta_i(\vec{p})|^2}{z-\epsilon_c^i(\vec{p}+\vec{w}_i)}\right)^{-1}.
\end{eqnarray}

\subsection{The Green's function of the conduction band}

Calculating the Green's function of the conduction band involves a treatment similar to that of the valence band. From equation (\ref{eqn_valGreendev}) and with definitions (\ref{eqn_anomGreenCc}) and (\ref{eqn_orderparam}) we get
\begin{eqnarray}\label{eqn_conductiontime}
\left(i\frac{\partial}{\partial t}-\epsilon_c^i(\vec{p}+\vec{w}_i)\right)&G_c^i(\vec{p},t,t')&=\delta(t-t')\nonumber\\&&-\Delta_i^*(\vec{p})F_i^\dagger(\vec{p},t,t'). 
\end{eqnarray}
The equation of motion of $F^\dagger$ is obtained with help of equation (\ref{eqn_MotionOps}) and Wick's theorem
\begin{eqnarray}\label{eqn_anomalrawCctime}
\lefteqn{\left(i\frac{\partial}{\partial t}-\epsilon_a(\vec{p})\right)F_i^\dagger(\vec{p},t,t')}\nonumber\\&&=-i\sum_{\vec{k},\vec{q},j}V_c(\vec{q})\langle Ta(\vec{p}+\vec{q},t)b_j^\dagger(\vec{k},t)b_j(\vec{k}-\vec{q},t)b_i^\dagger(\vec{p},t')\rangle\nonumber
\\&&\approx-i\sum_{\vec{q},j}V_c(\vec{q})\langle a(\vec{p}+\vec{q},t)b_j^\dagger(\vec{p}+\vec{q},t)\rangle\langle Tb_j(\vec{p},t)b_i^\dagger(\vec{p},t')\rangle.\nonumber\\
\end{eqnarray}
The averages on the right hand side bring into play three $b$ operators and present an off-diagonal term mixing $b_i$ with $b_j^\dagger$ operators. When $j=i$, the last average lets appear the Green's function $G_b^i$ while $i\neq j$ terms involve new Green's functions representing the scattering of an electron from one conduction band to another one (usually called multivalley scattering)
\begin{eqnarray}
H_{ij}(\vec{k},t,t')&=&(-i)\langle T b_i(\vec{k},t)b_j^\dagger(\vec{k},t')\rangle.\nonumber
\end{eqnarray}
Their equation of motion reads
\begin{eqnarray}\label{eqn_offdiagtime}
\left(i\frac{\partial}{\partial t}-\epsilon_c^i(\vec{p}+\vec{w}_i)\right)H_{ij}(\vec{k},t,t')=-\Delta_i^*(\vec{p})F_j^\dagger(\vec{p},t,t'). \end{eqnarray}
Thus, with the help of the definition of the order parameter $\Delta$, replacing this last definition into (\ref{eqn_anomalrawCctime}) results in
\begin{eqnarray}\label{eqn_anomalCctime}
\lefteqn{\left(i\frac{\partial}{\partial t}-\epsilon_a(\vec{p})\right)F_i^\dagger(\vec{p},t,t')}\nonumber\\&&=-\Delta_i(\vec{p})G_c^i(\vec{p},t,t')-\sum_{j\neq i}\Delta_j(\vec{p}) H_{ji}(\vec{p},t,t').
\end{eqnarray}
Equations (\ref{eqn_conductiontime}), (\ref{eqn_offdiagtime}) and (\ref{eqn_anomalCctime}) together build a system of equations which can be solved with respect to $G_c^i$, providing us with the following expression after a Fourier transform to frequency space
\begin{eqnarray} \label{eqn_conductionfreq}
\lefteqn{G_c^i(\vec{p}+\vec{w}_i,z)=\Bigg(z-\epsilon_c^i(\vec{p}+\vec{w}_i)}\nonumber\\&&\left.-\frac{|\Delta_i(\vec{p})|^2}{(z-\epsilon_v(\vec{p}))-\sum_{j\neq i}\frac{|\Delta_j(\vec{p})|^2}{z-\epsilon_c^j(\vec{p}+\vec{w}_j)}}\right)^{-1}.
\end{eqnarray}

\section{Results and discussions}\label{sec_results}

\subsection{The spectral function}\label{subsec_specfct}

In the context of photoemission, the spectral function $A(\vec{p},\Omega)$ plays a central role. It is directly proportional to the imaginary part of the Green's function and in the case of one-electron Green's functions as defined in section \ref{sec_model}, it describes the one-electron removal spectrum. 

For the excitonic condensate model, we distinguish the spectral function of the valence band
\begin{eqnarray} \label{eqn_valencespectfct}
A_v(\vec{p},\Omega)&=&-\frac{1}{\pi}\text{Im}[ G_v(\vec{p},\Omega+i\delta)]\nonumber
\end{eqnarray}
($\delta$ is here an infinitesimal real quantity) and that of the conduction band
\begin{eqnarray} \label{eqn_conductionspectfct}
A_c^i(\vec{p}+\vec{w}_i,\Omega)&=&-\frac{1}{\pi}\text{Im}[ G_c^i(\vec{p},\Omega+i\delta)].\nonumber
\end{eqnarray}
To simplify further calculations, we rewrite the Green's functions (equation (\ref{eqn_valencefreq}) and (\ref{eqn_conductionfreq})) in the following forms
\begin{eqnarray} 
G_v(\vec{p},z)&=&
\frac{1}{\mathcal{D}(\vec{p},z)}
\cdot\prod_i(z-\epsilon_c^i(\vec{p}+\vec{w}_i)),\label{eqn_valGreenDenom}\\
G_c^i(\vec{p},z)&=&\frac{1}{\mathcal{D}(\vec{p},z)}\cdot\Big((z-\epsilon_v(\vec{p}))\prod_{j\neq i}(z-\epsilon_c^j(\vec{p}+\vec{w}_j))\nonumber\\&&-\sum_{m,j\neq i}|\Delta_j(\vec{p})|^2|\varepsilon_{ijm}|(z-\epsilon_c^m(\vec{p}+\vec{w}_j))\Big)\nonumber\\\label{eqn_condGreenDenom}
\end{eqnarray}
($\varepsilon_{ijm}$ is the permutation symbol).
The denominator $\mathcal{D}$, common to all Green's functions, is
\begin{eqnarray}\label{eqn_denominator}
\mathcal{D}(\vec{p},z)&=&(z-\epsilon_v(\vec{p}))\prod_i(z-\epsilon_c^i(\vec{p}+\vec{w}_i))\nonumber\\&-&\sum_i|\Delta_i(\vec{p})|^2\prod_{j\neq i}(z-\epsilon_c^j(\vec{p}+\vec{w}_j))\nonumber\\
&=&\prod_{\alpha=1}^4(z-\Omega_\alpha(\vec{p}))
\end{eqnarray}
(here the index $\alpha$ refers to the four zeros of the denominator $\mathcal{D}$, while the other indices refer to the three conduction bands). In the last line, the denominator is factorized in terms involving its four (real) zeros $\Omega_\alpha(\vec{p})$ (which are implicitly functions of the order parameter $\Delta$). These zeros can be calculated exactly. However their analytical forms are too long to be written here. 

This allows us to break apart the Green's functions (\ref{eqn_valGreenDenom}) and (\ref{eqn_condGreenDenom}) into rational expressions with minimal denominators, so that we can use Sokhotsky's formula
\begin{eqnarray}
\frac{1}{x-x_0+i\epsilon}=\mathcal{P}\frac{1}{x-x_0}-i\pi\delta(x-x_0)\nonumber
\end{eqnarray}
($\mathcal{P}$ denotes the principal part) and write the spectral functions in rather simple forms
\begin{eqnarray} 
A_v(\vec{p},\Omega)&=&\sum_{\alpha=1}^4 P_\alpha^v(\vec{p})\:\delta(\Omega-\Omega_\alpha(\vec{p})),\nonumber\\
A_c^i(\vec{p}+\vec{w}_i,\Omega)&=&\sum_{\alpha=1}^4 P_\alpha^{c_i}(\vec{p})\:\delta(\Omega-\Omega_\alpha(\vec{p})),\nonumber
\end{eqnarray}
where the weights $P_\alpha(\vec{p})$ (which are also implicitly functions of the order parameter $\Delta$) associated to the poles $\Omega_\alpha$ are
\begin{eqnarray} 
P_\alpha^v(\vec{p})&=&\frac{\prod_i(\Omega_\alpha-\epsilon_c^i(\vec{p}+\vec{w}_i))}{\prod_{\beta\neq \alpha} (\Omega_\alpha-\Omega_\beta)},\nonumber\\
P_\alpha^{c_i}(\vec{p})&=&\frac{(\Omega_\alpha-\epsilon_v(\vec{p}))\prod_{n\neq i}(\Omega_\alpha-\epsilon_c^n(\vec{p}+\vec{w}_n))}{\prod_{\beta\neq \alpha} (\Omega_\alpha-\Omega_\beta)}\nonumber\\&&-\frac{\sum_{m,n\neq i}|\Delta_n(\vec{p})|^2|\epsilon_{inm}|(\Omega_\alpha-\epsilon_c^m(\vec{p}+\vec{w}_n))}{\prod_{\beta\neq \alpha} (\Omega_\alpha-\Omega_\beta)}.\nonumber
\end{eqnarray}
Until now, to ensure the generality of the theory, we always kept the $\vec{k}$-dependence of the order parameter $\Delta(\vec{k})$. However, in the practical analysis which will follow, we will use a $\vec{k}$-independent order parameter estimated from experiment.

\subsection{The spectral function and photoemission}\label{subsec_simulPES}
\begin{figure}
\centering
\includegraphics[width=8cm]{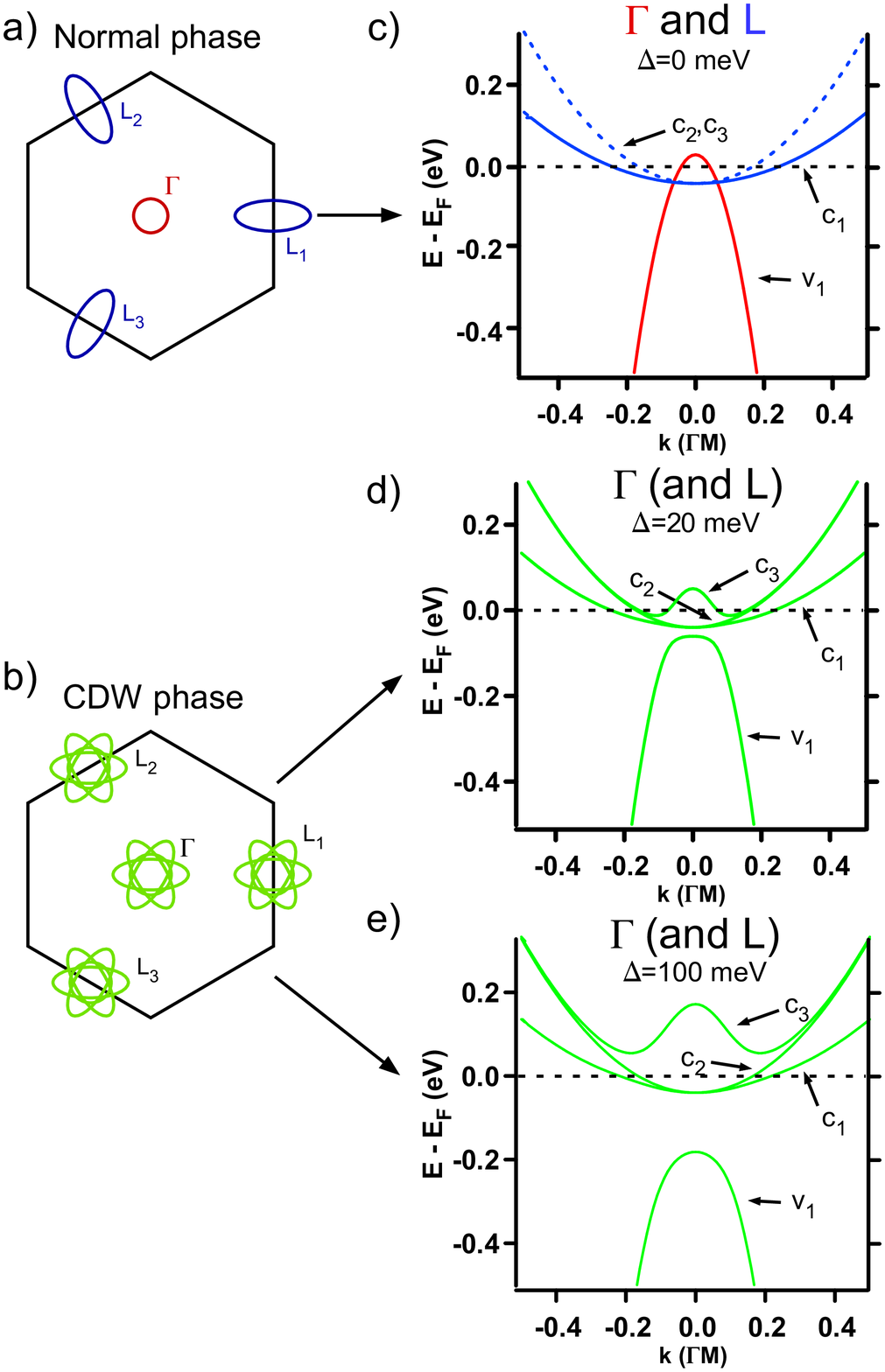}
\caption{\label{fig_CDWbands} (Color online) Schematic picture of the band positions in the model (near $E_F$) in the normal phase and in the CDW phase. Wave vectors are expressed in a multiple of $\Gamma M$. (a) In the normal phase ($\Delta=0$ meV), the Fermi surface composed of the valence band at $\Gamma$ (in red) and three symmetry equivalent conduction bands at $L$ (in blue). (b) In the CDW phase ($\Delta\neq0$ meV), $\Gamma$ becomes equivalent to $L$.  The electron pockets at $L$, backfolded to $\Gamma$, produce "flower"-like Fermi surfaces at each newly equivalent high symmetry point. (c) Dispersions calculated parallel to $\Gamma M$ (see Fig. 1(a)) around $\Gamma$ and parallel to $A L_1$ around the three $L$-points in the normal phase plotted on the same graph (the minima of the different conduction bands $c_1,c_2,c_3$ have been displaced from the $L$-points to $\Gamma$ on the graph). (d),(e) Dispersions around $\Gamma$ and along $\Gamma M$ in the CDW phase for $\Delta=20$ meV respectively $\Delta=100$. In the CDW phase, $\Gamma$ and $L$ become equivalent concerning the dispersions.}
\end{figure}
Within the sudden approximation, the contributions to the photoemission intensity are the spectral function, the matrix elements and the Fermi-Dirac distribution. In this paper we concentrate on the spectral function, being well established in the previous subsection.

We now choose to fix the parameters that describe the band dispersions. 
They will take the values
\footnote{The fit parameters are : $\epsilon_{v}^{0}$=-0.03$\pm$0.005 eV, $m_{v}$=-0.23$\pm$0.02 $m_{e}$,
where $m_{e}$ is the free electron mass, $t_{v}$=0.06$\pm$0.005 eV ; $\epsilon_{c}^{0}$=-0.01$\pm$0.0025 eV, $m_{c}^{x}$=5.5$\pm$0.2 $m_{e}$, $m_{c}^{y}$=2.2$\pm$0.1 $m_{e}$, $t_{c}$=0.03$\pm$0.0025 eV.}
 determined from our previous ARPES study\cite{CercellierPRL}.
From the spectral functions, we can derive a first important information. The zeros of their denominator $\mathcal{D}$ are the poles of the Green's functions and therefore represent the renormalized electronic band positions in the system. 
Noting that a zero order parameter $\Delta$ in equation (\ref{eqn_denominator}) results in band positions that are not renormalized, one realizes that $\Delta$ is a good indicator of the strength of this renormalization. Moreover, since this denominator is the same for both the valence and the conduction bands, we see that the valence band at $\Gamma$ is backfolded at $L$ and that the conduction bands at $L$ are backfolded at $\Gamma$ (as expected from the electron-hole coupling). This is a first indication of the CDW phase in the system. The situation for the band positions (not considering their spectral weight) is illustrated in Fig. \ref{fig_CDWbands}. Part (a) depicts a cut through the Fermi surface (FS) (around the $\Gamma$ and L points) in the normal phase, composed of the valence band hole pocket (red) and three (symmetry equivalent) electron pockets (blue). In the CDW phase, which is characterized by an order parameter $\Delta\neq0$ meV, this FS changes into that of Fig. \ref{fig_CDWbands} (b). Via the electron-hole coupling the $L$-points become equivalent to $\Gamma$ (not yet considering the spectral weights) and all three conduction bands are backfolded onto the valence band. 
In parallel, in Fig. \ref{fig_CDWbands} (c), (d) and (e) we show the associated dispersions. To facilitate the  comparison, in the normal phase, we superimpose the valence band ($v_1$ located at $\Gamma$, red continuous line) and the conduction bands (blue continuous line for $c_1$ and blue dashed line for $c_2,c_3$, normally located at the $L$-points but shifted here to $\Gamma$). For the CDW phase, we distinguish two cases, one with a low value of the order parameter (Fig. \ref{fig_CDWbands} (d), $\Delta=20$ meV) and one with a high value of the order parameter (Fig. \ref{fig_CDWbands} (e), $\Delta=100$ meV). These values are reasonable in comparison with experiment and help to understand how the CDW transition settles in. Once the order parameter increases to a non-zero value, there is a strong change in the band dispersions. The valence band $v_1$ and the conduction band $c_3$ split, opening a gap between them. As the order parameter increases to $\Delta=100$ meV, $v_1$ and $c_3$ repell each other further, while $c_1$ and $c_2$ stay at their original positions. 

\begin{figure}
\centering
\includegraphics[width=8cm]{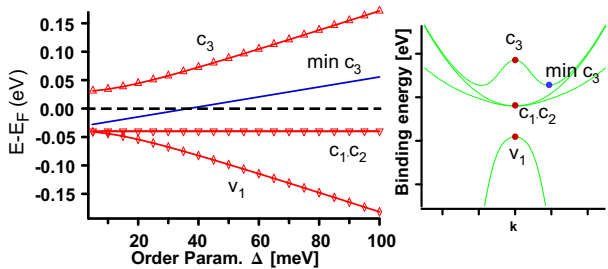}
\caption{\label{fig_bandspos} (Color online) Extrema of the renormalized bands as a function of the order parameter $\Delta$ (left). Position of these extrema on the band dispersions (right).}
\end{figure}

We now turn to the discussion of the extrema of the bands. Fig. \ref{fig_bandspos} presents the band extrema as a function of the order parameter $\Delta$. We see that, except for small values of the order parameter, they display a linear behaviour (this can be shown analytically from the denominator in equation (\ref{eqn_denominator}) for $\vec{k}$ exactly at $\Gamma$ and $L$ where the three original conduction bands have the same energy). Among the three conduction bands, only one ($c_3$) shifts away from the Fermi energy. Its two minima (blue line) also increase linearly. Compared to the conduction band $c_3$, the valence band maximum ($v_1$) follows the inverse behaviour, thereby opening a gap below $E_F$.

It is important to realize that when considering three conduction bands instead of one, the system remains in a semimetallic state at low temperature rather than evolving into an insulating state, since the gap opens below the Fermi energy. Therefore, strictly speaking, the denomination "excitonic insulator phase" in this context is misleading and we rather adopted the expression "exciton condensate phase".

\begin{figure}
\centering
\includegraphics[width=8.5cm]{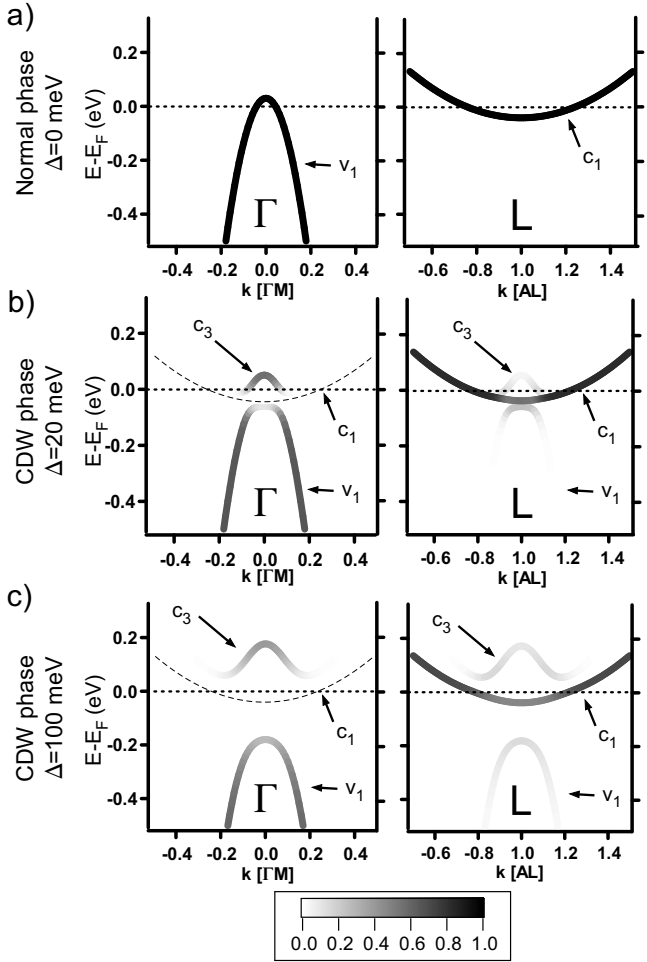}
\caption{\label{fig_bandsspec} Band dispersions with their corresponding spectral weight at $\Gamma$ and $L$, along $\Gamma M$ and $AL$ directions, respectively. Graph (a) describe the normal phase ($\Delta=0$ meV), (b) the CDW phase with moderate excitonic effects ($\Delta=20$ meV) and (c) the CDW phase with strong excitonic effects ($\Delta=100$ meV). The dashed lines indicate a band ($c_1$) having a small non-zero SW (see text).
}
\end{figure}

Besides the position of the bands the spectral function contains an additional crucial information, namely, the spectral weight (SW) carried by each band in the process of one-electron removal probed by photoemission. It is related to the numerator of the spectral function. We now add this feature to the previous figure and obtain Fig. \ref{fig_bandsspec}, where the SW of the bands is indicated in gray scale.
The evolution from the normal state (Fig. \ref{fig_bandsspec}(a)) to the CDW state with an order parameter of 20 meV (Fig. \ref{fig_bandsspec}(b)) and 100 meV (Fig. \ref{fig_bandsspec}(c)) is shown. 

\begin{figure}
\centering
\includegraphics[width=8cm]{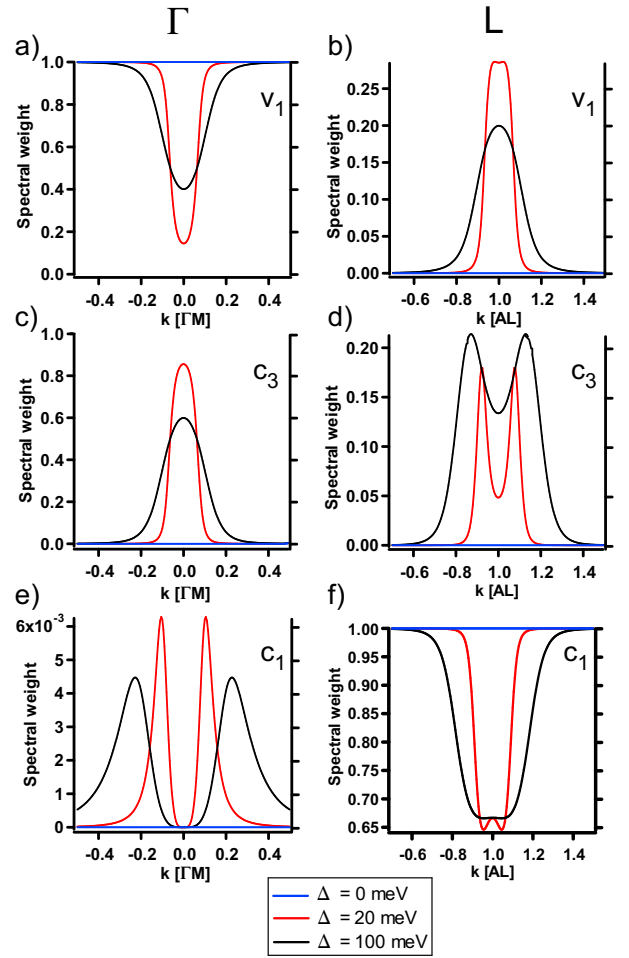}
\caption{\label{fig_specweights} (Color online) Spectral weights of the bands at $\Gamma$ and $L$, along $\Gamma M$ and $AL$ directions, respectively, for $\Delta$ values of 0, 20 and 100 meV. Graphs (a) and (b) describe the valence band ($v_1$) at $\Gamma$ and its backfolded version at $L$, respectively. Graphs (c) and (d) describe the conduction band $c_3$ at $\Gamma$ (where it follows the top of the original valence band) and $L$ respectively. Graphs (e) and (f) describe the conduction band $c_1$ at $\Gamma$ and $L$, respectively (where it is the original conduction band).
}
\end{figure}

In Fig. \ref{fig_specweights}, we focus on the SW of the bands at $\Gamma$ and $L$ 
(SW of the conduction band $c_2$ is not represented here since it is exactly $0$ for every $k$ along the $\Gamma M$ and $AL$ directions). 
Graphs \ref{fig_specweights}(a) and (b) display the SW of the valence band ($v_1$) at $\Gamma$ and $L$, respectively, for different values of the order parameter $\Delta$. Graphs  \ref{fig_specweights}(c), (d) and \ref{fig_specweights}(e), (f) show the SW for bands $c_3$ and $c_1$, respectively.

We immediately see (Fig. \ref{fig_bandsspec}) that, with respect to the SW, the backfolding is in fact incomplete even at a large value of the order parameter. Indeed, in the CDW phase with an order parameter of $\Delta=100$ meV (Fig.  \ref{fig_bandsspec}(c)), at $\Gamma$, the original valence band conserves $40\%$ of its (normal phase) SW (Fig.  \ref{fig_specweights}(a)), while the backfolded conduction band ($c_3$) carries a SW of $60\%$ (Fig.  \ref{fig_specweights}(c)). At $L$, the situation is more complicated, since three bands ($v_1$, $c_1$, $c_3$) share now the SW. The original conduction band ($c_1$) keeps a minimum of $67\%$ of SW (Fig.  \ref{fig_specweights}(f)), while the other two backfolded bands, namely a symmetry equivalent conduction band ($c_3$) and the valence band, divide among themselves the remaining $33\%$ (Fig.  \ref{fig_specweights}(b), (d)).

We also present the graphs for $\Delta=20$ meV. 
We see that there is a large SW loss (more than $80\%$ at $\vec{k}=\vec{0}$ \AA$^{-1}$) in the valence band, even larger than that for $\Delta=100$ meV (Fig.  \ref{fig_specweights}(a)). What happens can be seen as follows. If we observe the graphs in Figs. \ref{fig_bandsspec} (a), (b) and (c) close to $\Gamma$ (i.e., the left panels) as a function of $\Delta$, we see that at $\Delta = 20$ meV the shape of the valence band tries to stay the same as for
 $\Delta = 0$ meV. This is achieved by a reduced SW of $v_1$ and an increased SW of $c_3$. At  $\Delta = 100$ meV the dispersions are sufficiently different from the non-renormalized ones to carry more SW. 
In other words, for small values of the order parameter, the SW will be distributed along the parts of the dispersions corresponding mainly to the non-renormalized ones. 
A similar situation happens at $L$, as can be seen in Fig. \ref{fig_bandsspec} between bands $v_1$, $c_1$ and $c_3$. But this time, the original band (in the normal phase) is the conduction band $c_1$, so that for small values of the order parameter, its SW is shared between $v_1$, $c_1$ and $c_3$.
In Fig. \ref{fig_bandsspec}, the dashed lines indicate a conduction band ($c_1$) backfolded to $\Gamma$, which has a small non-zero SW as shown in Graph \ref{fig_specweights}(e) (it is less than $1\%$ for the values of the order parameter considered here).

It should be noted that from photoemission data it is difficult to extract information concerning (thermally occupied) states above the Fermi energy (set to $0$ eV here), so that SW of band $c_3$ is hardly measured in experiment\cite{CercellierPRL}.

\subsection{Comparison with experiment}\label{subsec_experiment}

To emphasize the good agreement between our model and experiment, we further analyze experimental ARPES intensity maps\cite{CercellierPRL} in the light of the previous discussions (calculated intensity maps are not reproduced hereafter, see\cite{CercellierPRL} for more details). The data was collected at the Swiss Light Source with a photon energy of 31 eV on TiSe$_2$ samples (at this photon energy, the normal emission spectra correspond to states located close to the $\Gamma$ point).

\begin{figure}
\centering
\includegraphics[width=8cm]{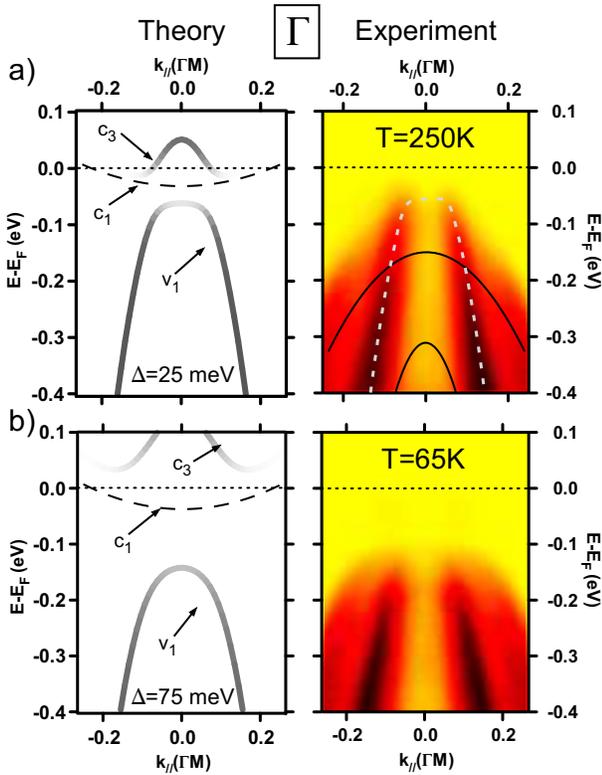}
\caption{\label{fig_measG} (Color online) Comparisons between theoretical and experimental ($h\nu=31$ eV) electronic structures at $\Gamma$. (a) The theoretical bands have been calculated for $\Delta=25$ meV and the experimental intensity maps are taken at $T=250$K. The continuous black lines highlight the Se$4p$-derived bands not considered in the model, while the dashed white line indicates the valence band corresponding to $v_1$. (b) The theoretical bands have been calculated for $\Delta=75$ meV and the experimental intensity maps are taken at $T=65$K. The dashed black lines indicate the backfolded conduction band $c_1$ which carries a small non-zero SW (Fig. \ref{fig_specweights} (e)).
}
\end{figure}

Fig. \ref{fig_measG} presents comparisons between the theoretical (left) and experimental (right) electronic structures at $\Gamma$. The experimental intensity maps at $T=250$K (Fig. \ref{fig_measG} (a)) and $T=65$K (Fig. \ref{fig_measG} (b)) are compared to calculated bands with $\Delta=25$ meV and $\Delta=75$ meV, respectively. At $T=250$K on the experimental side (Fig. \ref{fig_measG}, right), the situation is more complicated than in our model. Indeed, there are three Se$4p$-derived valence bands, out of which two (black lines), are not considered in our model. The dashed white line corresponds to the valence band $v_1$ of the model, which suffers already SW loss at $T=250$K. It flattens at its maximum and deviates from the parabolic shape of the normal phase dispersion (this is clear from an energy distribution curve taken at $k_\parallel=0$ \AA$^{-1}$, not shown here). On the theoretical side (Fig. \ref{fig_measG} (a), left), the dispersion reproduces well the experiment when considering an order parameter of $\Delta=25$ meV. However, at this temperature, the system should be in the normal phase. 
Nevertheless, as in high temperature superconductors, above the critical temperature, we expect fluctuations to persist well above $T_c$ in the excitonic condensate phase\cite{acoustics,XRDWoo}. Thus, this non-zero order parameter above $T_c$ may be understood in terms of fluctuations. According to the left graph of Fig. \ref{fig_measG} (a), the bottom of the backfolded conduction band $c_3$ appears just below $E_F$. In parallel, the experiment shows small humps in momentum distribution curves near $E_F$ (not seen in the false colour map here). These can be attributed to $c_3$, considering that the Fermi distribution will weaken the SW of this band on the theoretical dispersion. 
At $T=65$K (Fig. \ref{fig_measG} (b), right), the valence band in the experimental intensity map shifts to higher binding energies, in agreement with the theoretical dispersions calculated for an order parameter $\Delta=75$ meV (Fig. \ref{fig_measG} (b), left). Moreover, on the experimental map, some intensity emerges just below $E_F$ (not seen in the false colour map, see Ref.\cite{CercellierPRL}), revealing a dispersive band. Altough it does not appear directly in the corresponding calculation using $\delta$-peaks, it is reproduced if a finite 30 meV line broadening (lifetime) is introduced. In other words, this dispersive intensity comes from the combined tails of the maximum of the valence band $v_1$ (located in the occupied states) and of the minima of the backfolded conduction band $c_3$ (located in the unoccupied states).

\begin{figure}
\centering
\includegraphics[width=8cm]{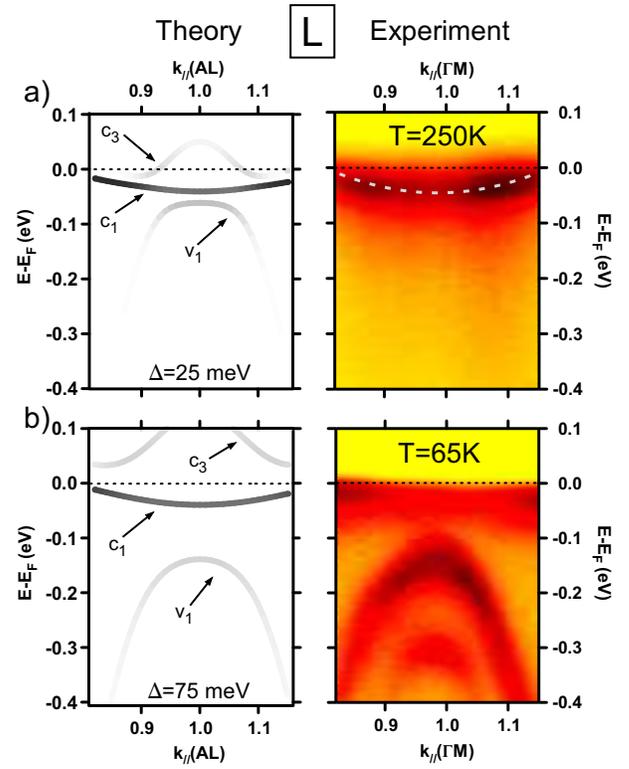}
\caption{\label{fig_measM} (Color online) Comparisons between theoretical and experimental ($h\nu=31$ eV) electronic structures at the Brillouin zone boundary (see text). (a) The theoretical bands have been calculated for $\Delta=25$ meV and the experimental intensity maps are taken at $T=250$K. The dashed white line indicates the conduction band corresponding to $c_1$. (b) The theoretical bands have been calculated for $\Delta=75$ meV and the experimental intensity maps are taken at $T=65$K.
}
\end{figure}

Fig. \ref{fig_measM} presents comparisons between the theoretical and experimental electronic structures at the Brillouin zone boundary. The experimental intensity maps at $T=250$K (Fig. \ref{fig_measM} (a)) and $T=65$K (Fig. \ref{fig_measM} (b)) are compared to calculated bands with $\Delta=25$ meV and $\Delta=75$ meV, respectively. At a photon energy of 31 eV, initial states between $M$ and $L$ are probed (in a free electron final state picture, we are not probing the BZ exactly at $L$). For simplicity, we will continue using the $L$ notation below. Substantial changes (explained by excitonic effects) in the spectra are evident.
At $T=250$K (Fig. \ref{fig_measM} (a)), on the experimental side (right), the conduction band, highlighted by the dashed white line, is well described by the theoretical band $c_1$ (left). It also exhibits a decrease of intensity at its bottom, as predicted by our model (see Fig. \ref{fig_specweights} (f)). Moreover it looks like a band having a large width and some surprisingly high intensity persists far from its centroid, even at binding energies of about 300 meV. These features can be well explained by the theoretical bands (Fig. \ref{fig_measM}, left) at $\Delta=25$ meV. Due to the finite width of real bands, the conduction band $c_1$ merges with its close (backfolded) neighbours $v_1$ or $c_3$ (depending on the position along $AL$), resulting in a band broadening. The residual intensity at high binding energies is explained with the populated branches of the backfolded valence band $v_1$ (see Fig. \ref{fig_specweights} (b)). Indeed, looking carefully at the lower part of the experimental map (Fig. \ref{fig_measM} (a), right), one sees that the residual intensity is larger away from $L$ (this is confirmed by momentum distribution curves, not shown here).
At $T=65$K (Fig. \ref{fig_measM} (b)), on the experimental side (right), a strong signature of the CDW appears. The valence band is backfolded at $L$ with a high SW. In fact, even a second Se$4p$-derived valence band (indicated by the lowest lying black line in the right graph of Fig. \ref{fig_measG} (a)) participates to the backfolding at $L$. On the theoretical side, this situation (considering only the topmost Se$4p$-derived valence band) is reproduced with an order parameter of $\Delta=75$ meV. The backfolded valence band $v_1$ is well separated from the conduction band $c_1$, as in the experiment. It can be clearly seen in the corresponding experimental intensity map that the conduction band $c_1$ does not shift with an increasing order parameter (see Fig. \ref{fig_bandspos}) and looses more SW at its bottom (see Fig. \ref{fig_specweights} (f)). The backfolded conduction band $c_3$ is too far away from $E_F$ in the unoccupied states to be measured by ARPES. In the model, at $L$, the intensity of the backfolded valence band $v_1$ is lower than that of the conduction band $c_1$ for high values of the order parameter, corresponding to a well settled CDW phase (see Fig. \ref{fig_specweights} (b) and (f)). In the ARPES measurements presented here, this intensity relation is reversed, as can be seen on the right graph of Fig. \ref{fig_measM} (b). The precise reason for this matter remains unclear. It can be due to the fact that we consider only the topmost valence band in our model. Indeed, a second backfolded valence band appears at $L$ in the low temperature measurements (Fig. \ref{fig_measM} (b), right). Moreover, we have noticed that this intensity relation between the original and backfolded bands can change from one sample to another, or even depends on the quality of the cleaved surface. Further investigations are needed to understand this issue. 

\subsection{Further discussions}\label{subsec_discussions}

In the model described in section \ref{sec_model}, the chemical potential was not explicitly calculated since it was defined as the zero energy of the dispersions. To verify whether the chemical potential shifts when the system enters in the CDW phase, we have computed the electronic density for the renormalized bands, taking into account their dispersion over the whole BZ. Due to the parabolic approximation of the band dispersions around their extrema, we only took into account electrons having an energy up to 0.5 eV below $E_F$. If we keep the chemical potential at $\mu=0$ eV, a decrease of about $35\%$ of the electronic density results between the normal phase and the CDW phase with an order parameter $\Delta=75$ meV. This discrepancy is reduced to zero if we shift the chemical potential by $~+60$ meV. This result can be understood with the SW transfers depicted on Fig. \ref{fig_bandsspec}. At $\Gamma$, when going from the normal to the CDW phase (from Fig. \ref{fig_bandsspec} (a) to (c)) we loose $13\%$ of the SW of the normal phase (integration of the SWs of Fig. \ref{fig_specweights} (a) and (c)) accounting for the Fermi distribution. At $L$ (from Fig. \ref{fig_bandsspec} (a) to (c)), after the CDW transition, we loose $40\%$ of SW in the conduction band and we acquire $18\%$ of SW in the backfolded valence band (integration of the SWs of Fig. \ref{fig_specweights} (b), (d) and (f)). Thus, considering only the high symmetry directions for illustrative purposes, this results in the $35\%$ of SW missing when going from the normal to the CDW phase, which can be recovered by slightly raising the chemical potential (which affects mostly the conduction band $c_1$).

This shift would be measurable in photoemission. For having a detailed knowledge of the chemical potential, one needs to perform precise ARPES measurements over a wide range of temperatures. However, this is beyond the scope of this article and will be studied in the future.
\\

From previous ARPES data\cite{CercellierPRL}, we extracted an overlap between the valence and conduction bands of $30$ meV. Density functional theory calculation for 1\textit{T}-TiSe$_2$ within DFT\cite{FangLDA} predicts a semimetallic system with an overlap of about $800$ meV. However this is not completely incompatible with our measurements. Indeed, it appears that at room temperature excitonic effects are already present and the valence band is shifted to higher binding energies, below the minimum of the conduction band (as we see in Fig. \ref{fig_bandspos} for $\Delta\neq 0$ meV). It must be emphasized that this will be the case, irrelevant of the position of the valence band, as long as the valence band maximum is above the conduction band minimum in the normal phase. In other words, switching on excitonic effects in a system having in the normal phase a bandstructure similar to that predicted by DFT would produce room temperature dispersions very similar to what we measure.

\section{Conclusions}\label{sec_conclusion}

Recently intensity maps calculated within the excitonic condensate phase model have been compared with ARPES data of 1\textit{T}-TiSe$_2$. Strikingly good agreement gave strong evidence for excitonic condensation as the driving force of the CDW transition\cite{CercellierPRL}. In the present paper, we have presented the theory of the excitonic insulator model generalized to the three dimensional case of 1\textit{T}-TiSe$_2$ with anisotropic band dispersion. From the Green's functions of the valence and conduction bands, we computed the corresponding spectral functions needed to generate photoemission intensity maps. The mathematical treatment is similar to BCS theory and the deduced order parameter in the low temperature phase describes the intensity of condensating electron-hole pairs (excitons). These pairs are created by the electron-hole interaction between the valence band at $\Gamma$ and the conduction bands at $L$. As a natural consequence of the non-zero momentum of the excitons, this produces band backfoldings between $\Gamma$ and $L$ which thus tend to be equivalent as the order parameter increases. It must be emphasized that the CDW produced by this model is of purely electronic origin and that the spectral weights transferred between the original and backfolded bands are large (see reference\cite{CercellierPRL} for a more complete discussion of this subject). While no real gap opens at the Fermi energy, it is notably shown that the valence band (original and backfolded) is shifted in a quasi-linear manner to higher binding energies as the order parameter increases. Such a behaviour could offer a direct way to extract the temperature dependence of the excitonic order parameter. Finally, the present paper treats only the low temperature condensation phase of the excitonic condensate. However, room temperature measurements indicate that strong excitonic fluctuations prevail far above $T_c$. Their theoretical and experimental study promises an interesting extension of this work.

\begin{acknowledgments}
Skillfull technical assistance was provided by the workshop and electric engineering team. This work was supported by the Fonds National Suisse pour la Recherche Scientifique through Div. II and MaNEP.
\end{acknowledgments}

\bibliography{TiSe2_theo_vf}
\end{document}